\documentclass[12pt]{article}

\begin{document}

\begin{titlepage}

\begin{flushright}
CCTP-10-09 \\ UFIFT-QG-10-07
\end{flushright}

\begin{center}
{\bf Stochastic Samples versus Vacuum Expectation Values in Cosmology}
\end{center}

\begin{center}
N. C. Tsamis$^{\dagger}$ and Aggelos Tzetzias$^{\ddagger}$
\end{center}

\begin{center}
\it{Department of Physics, University of Crete \\
GR-710 03 Heraklion, HELLAS}
\end{center}

\begin{center}
R. P. Woodard$^{\ast}$
\end{center}

\begin{center}
\it{Department of Physics, University of Florida \\
Gainesville, FL 32611, UNITED STATES}
\end{center}

\vspace{1cm}

\begin{center}
ABSTRACT
\end{center}
Particle theorists typically use expectation values to study the
quantum back-reaction on inflation, whereas many cosmologists stress 
the stochastic nature of the process. While expectation values certainly 
give misleading results for some things, such as the stress tensor, we 
argue that operators exist for which there is no essential problem. We 
quantify this by examining the stochastic properties of a noninteracting, 
massless, minimally coupled scalar on a locally de Sitter background. The 
square of the stochastic realization of this field seems to provide an
example of great relevance for which expectation values are not misleading.
We also examine the frequently expressed concern that significant 
back-reaction from expectation values necessarily implies large stochastic 
fluctuations between nearby spatial points. Rather than viewing the 
stochastic formalism in opposition to expectation values, we argue that it 
provides a marvelously simple way of capturing the leading infrared 
logarithm corrections to the latter, as advocated by Starobinsky.

\begin{flushleft}
$^{\dagger}$ e-mail: tsamis@physics.uoc.gr \\
$^{\ddagger}$ e-mail: tzetzias@physics.uoc.gr \\
$^{\ast}$ e-mail: woodard@phys.ufl.edu
\end{flushleft}

\end{titlepage}

\section{Introduction}

Schr\"odinger was the first to suggest that spacetime expansion can 
lead to particle production by ripping virtual particles out of the 
vacuum \cite{Schr}. Following early work by Imamura \cite{TI}, the 
first quantitative results were obtained by Parker \cite{Parker1}. He 
found that the effect is maximized during accelerated expansion, and for 
massless particles which are not conformally invariant \cite{Parker2},
such as massless, minimally coupled scalars and (as noted by Grishchuk 
\cite{Grishchuk}) gravitons. Precisely this process is responsible for the 
primordial spectra of scalar and tensor perturbations which are believed 
to arise from inflation \cite{SMC}, the scalar contribution of which
has been imaged \cite{WMAP}.

Inflationary particle production results from the background gravitational 
field acting on quantum matter (and graviton) fluctuations. It is natural 
to wonder about the complementary process of {\it back-reaction} in which
the newly produced particles modify the background gravitational field, 
either directly or through their self-interactions. People who approach 
this from the perspective of particle physics typically attempt to quantify 
back-reaction using expectation values or in-out matrix elements 
\cite{Polyakov,Myhrvold,Habib,Ford,Mottola,Tomaras,Us,Brandenberger,Abramo1,
Abramo2,Parker3,Veneziano}.

Many cosmologists dismiss the use of expectation values and in-out matrix
elements as giving an unreliable average over vastly different portions of 
a quantum wave function which has actually decohered.\footnote{For an
excellent recent study of cosmological decoherence, which does not 
necessarily endorse the anti-VEV position, see \cite{decoh}.} They fear 
that particle theorists are falling victim to a sort of cosmological
Schr\"odinger Cat Paradox based on a fictitious, mean geometry which bears 
no relation to what any observer would experience. Cosmologists prefer to 
instead study back-reaction using a stochastic formalism in which the 
super-horizon modes of various fields are regarded as classical, random 
variables \cite{AAS,AVNS,Linde,RST,SJRSNNWVMMENPR}. Whereas expectation 
values in a homogeneous and isotropic state necessarily produce a 
homogeneous and isotropic geometry, cosmologists assert that the actual
universe is not even approximately homogeneous on super-horizon scales 
\cite{Linde}. They also fear that if the back-reaction inferred from 
expectation values ever becomes significant then the resulting universe 
would show unacceptable spatial fluctuation.

There is no question that cosmologists are right about certain operators
being poorly described by their expectation values. For example, the
vacuum expectation value of the stress-energy tensor is homogeneous and 
isotropic, whereas we perceive inhomogeneities and anisotropies. What isn't 
clear is whether or not expectation values are in all cases misleading. 
Such an extreme position would be embarrassing for cosmologists (in spite 
of the fact that some do advocate it) because the primordial power spectra 
are defined by taking expectation values \cite{reviews}.

We suspect that the reliability of expectation values depends upon
the operator under study. For operators which average to zero, such
as the density perturbation, the entire result arises from the
decoherence effect, so one makes an enormous mistake by ignoring it.
Other operators --- for example, the square of a scalar field ---
acquire a significant homogeneous expectation value upon which spatial
variations are superimposed. Any quantum fluctuation drives this
sort of operator positive, so one might happen to inhabit a special
region of the universe in which there is little effect for a long
time, but there will sooner or later be a large effect. The
expectation value of such an operator can correctly reflect the
long-term trend everywhere in space, even though it misses variations
from one region to another.

The purpose of this paper is to use the noninteracting, massless,
minimally coupled scalar on de Sitter background to give a quantitative 
assessment of the two key issues under dispute between particle theorists 
and cosmologists:
\begin{enumerate}
\item{How unreliable are expectation values? and}
\item{How much spatial variation should one expect?}
\end{enumerate}
In section 2 we review the de Sitter geometry, the scalar field model,
and our stochastic realization of it. In section 3 we demonstrate that 
the square of the stochastic field follows a $\chi^2$ distribution whose 
mean grows with the number of e-foldings. Although fluctuations about 
this mean are significant, they do not contradict the picture provided 
by expectation values. In section 4 we address the issue of spatial 
fluctuations by studying the difference of the stochastic scalar at two 
points held at a fixed physical distance from one another. Our 
conclusions comprise section 5.

\section{The Theoretical Context}

We work on the spacetime manifold $T^3 \times R$. We assume each
of the three spatial coordinates lies in the range, $0 \leq x^i < H^{-1}$, 
where $H$ is the Hubble constant. We also assume that the geometry is 
locally de Sitter,
\begin{equation}
ds^2 = -dt^2 + a^2(t) d\vec{x} \cdot d\vec{x} \qquad , \qquad
a(t) \equiv e^{H t} \; . \label{deS}
\end{equation}
Note that the scale factor $a(t)$ is normalized to unity at $t = 0$, 
rather than at the current time. Because the geometry (\ref{deS}) is 
homogeneous and isotropic we can expand any function of spacetime in
a spatial Fourier series,
\begin{equation}
f(t,\vec{x}) = \sum_{\vec{n}} f_{\vec{n}}(t) e^{i\vec{k} \cdot \vec{x}}
\qquad , \qquad \vec{k} = 2 \pi H \vec{n} \; .
\end{equation}
A key event for any mode of wave number $k \equiv \Vert \vec{k} \Vert$
is the time $t_k$ of horizon crossing when its physical wave number
redshifts down to the Hubble constant,
\begin{equation}
\frac{k}{a(t_k)} = H \; .
\end{equation}

The Lagrangian for a massless, minimally coupled scalar is,
\begin{equation}
\mathcal{L} = -\frac12 \partial_{\mu} \varphi \partial_{\nu} \varphi
g^{\mu\nu} \sqrt{-g} = \frac12 a^3 \dot{\varphi}^2 - \frac12 a \Vert 
\vec{\nabla} \varphi \Vert^2 \; .
\end{equation}
For all nonzero wave numbers $k$ the canonically normalized, Bunch-Davies
mode functions are,
\begin{equation}
u(t,k) = \frac{H}{\sqrt{2 k^3}} \Bigl[1 - \frac{i k}{H a(t)} \Bigr]
\exp\Bigl[ \frac{i k}{H a(t)} \Bigr] \qquad \Longrightarrow \qquad
u \dot{u}^* - \dot{u} u^* = \frac{i}{a^3} \; .
\end{equation}
For the $k = 0$ mode the two solutions are a constant and $1/a^3(t)$.
Hence the field operator can be expanded as,
\begin{equation}
\varphi(t,\vec{x}) = H^{\frac32} \Biggl\{ Q - \frac{P}{3 H a^3(t)}
+ \sum_{\vec{n} \neq 0} \Biggl[ u(t,k) e^{i \vec{k} \cdot \vec{x}} 
A_{\vec{n}} + u^*(t,k) e^{-i \vec{k} \cdot \vec{x}} A^{\dagger}_{\vec{n}}
\Biggr] \Biggr\} \; , \label{freefield}
\end{equation}
where the nonvanishing commutators are,
\begin{equation}
\Bigl[Q, P\Bigr] = i \qquad , \qquad \Bigl[A_{\vec{m}} , 
A^{\dagger}_{\vec{n}} \Bigr] = \delta_{\vec{m} , \vec{n}} \; .
\end{equation}
The state $\vert \Omega \rangle$ which is annihilated by all $A_{\vec{n}}$
is known as Bunch-Davies vacuum. It does not matter very much what we
assume about its dependence upon the $0$-mode coordinate.

The quantity $\varphi(t,\vec{x})$ is a quantum field operator which obeys
the Uncertainty Principle,
\begin{equation}
\Bigl[ \varphi(t,\vec{x}) , \dot{\varphi}(t,\vec{y}) \Bigr] = 
i \delta^3(\vec{x} \!-\! \vec{y}) \; .
\end{equation}
Because the free field expansion (\ref{freefield}) contains arbitrarily 
large wave numbers, expectation values of coincident products of 
$\varphi(t,\vec{x})$ can harbor ultraviolet divergences. All of these 
features are absent in the stochastic realization of $\varphi(t,\vec{x})$
which we construct by taking the infrared limit of the mode functions,
\begin{equation}
\lim_{ k \ll H a} u(t,k) = \frac{H}{\sqrt{2 k^3}} \; ,
\end{equation}
and retaining only the super-horizon modes. We denote this stochastic 
field as $\widehat{\varphi}(t,\vec{x})$ and its definition is \cite{Abramo2},
\begin{equation}
\widehat{\varphi}(t,\vec{x}) \equiv \sum_{\vec{n} \neq 0} 
\sqrt{\frac{H^5}{2 k^3}} \, \theta\Bigl( H a(t) \!-\! k\Bigr) 
\Bigl[ e^{i \vec{k} \cdot \vec{x}} \widehat{A}_{\vec{n}} +
e^{-i \vec{k} \cdot \vec{x}} \widehat{A}^*_{\vec{n}} \Bigr] \; .
\end{equation}
Instead of being creation and annihilation operators, $\widehat{A}_{\vec{n}}$ 
and $\widehat{A}^*_{\vec{n}}$ are considered to be (complex conjugate) 
stochastic random variables which follow a normal distribution with mean 
zero and standard deviation one. That is, one can express 
$\widehat{A}_{\vec{n}}$ as the sum of two real, independent random variables,
\begin{equation}
\widehat{A}_{\vec{n}} \equiv \alpha_{\vec{n}} + i \beta_{\vec{n}} \qquad 
{\rm where} \qquad \rho\Bigl(\alpha_{\vec{n}} = x,\beta_{\vec{n}} = y\Bigr) 
= \frac1{2\pi} e^{-\frac12 (x^2 + y^2)} \; . \label{real}
\end{equation}

\section{Fluctuations about the Mean}

It is useful to express $\widehat{\varphi}(t,\vec{x})$ in terms of the
real stochastic variables $\alpha_{\vec{n}}$ and $\beta_{\vec{n}}$ which
were introduced in (\ref{real}),
\begin{equation}
\widehat{\varphi}(t,\vec{x}) = \sum_{\vec{n} \neq 0} \sqrt{\frac{H^5}{2 k^3}}
\, \theta\Bigl( H a(t) \!-\! k\Bigr) \Bigl[ \cos(\vec{k} \!\cdot\! \vec{x})
\, \alpha_{\vec{n}} - \sin(\vec{k} \!\cdot\! \vec{x}) \, \beta_{\vec{n}} 
\Bigr] \; .
\end{equation}
In this form one can recognize $\widehat{\varphi}(t,\vec{x})$ as the sum 
of a vast number $N(t)$ of independent Gaussian random variables,
\begin{equation}
N(t) = 2 \sum_{\vec{n} \neq 0} \theta\Bigl( H a(t) \!-\! k\Bigr) 
\simeq 2 \times 4\pi \int_{1/2\pi}^{a/2\pi} \!\!\!\!\!\!\!\!\!\! dn \, n^2 
\simeq \frac{a^3(t)}{3 \pi^2} \; .
\end{equation}
Of course the sum of any number of independent Gaussian random variables
gives another Gaussian random variable whose mean is the sum of the means
and whose variance is the sum of the variances. Because the mean of each
variable is zero, the mean of $\widehat{\varphi}(t,\vec{x})$ vanishes. On
the other hand, its variance grows,
\begin{eqnarray}
\sigma^2(t) & = & \sum_{\vec{n} \neq 0} \frac{H^5}{2 k^3}
\, \theta\Bigl( H a(t) \!-\! k\Bigr) \Bigl[ \cos^2(\vec{k} \!\cdot\! \vec{x})
+ \sin^2(\vec{k} \!\cdot\! \vec{x}) \Bigr] \; , \\
& \simeq & 4\pi \int_{1/2\pi}^{a/2\pi} \!\!\!\!\!\!\!\!\!\! dn \, n^2 \,
\frac{H^5}{2 (2\pi H n)^3} = \frac{H^2}{4 \pi^2} \, \ln[a(t)] \; . \label{sig1}
\end{eqnarray}
Hence we can say that, at any given spacetime point, the stochastic field 
$\widehat{\varphi}(t,\vec{x})$ follows a normal distribution with mean zero
and variance $\sigma^2(t)$,
\begin{equation}
\rho\Bigl( \widehat{\varphi}(t,\vec{x}) = Z\Bigr) = \frac1{\sqrt{ 2 \pi 
\sigma^2(t)}} \, \exp\Bigl[ -\frac{Z^2}{2 \sigma^2(t)} \Bigr] \; .
\end{equation}
The fields at different spacetime points are not statistically independent, 
but we will not need to worry about that until the next section.

The variable $\widehat{\varphi}(t,\vec{x})$ provides a classic example
of why cosmologists distrust expectation values. Although its expectation
value vanishes, the sto\-chas\-tic field experiences very significant and 
growing fluctuations, as its variance reveals. Someone interested in the 
behavior of $\widehat{\varphi}(t,\vec{x})$ would derive a completely 
misleading picture from its expectation value.

On the other hand, consider the variable $\widehat{\varphi}^2(t,\vec{x})$. 
Because it is the square of a Gaussian with mean zero it follows a
$\chi^2$ distribution whose mean is the variance of the original Gaussian,
\begin{equation}
\rho\Bigl( \widehat{\varphi}^2(t,\vec{x}) = Z\Bigr) = \frac1{\sqrt{2 \pi 
\sigma^2(t) Z }} \, \exp\Bigl[ -\frac{Z}{2 \sigma^2(t)} \Bigr] \; . 
\label{chisq}
\end{equation}
The variance of $\widehat{\varphi}^2(t,\vec{x})$ might seem to vindicate
the extreme cosmologist position that expectation values are never
reliable,
\begin{equation}
\Biggl\langle \Bigl( \widehat{\varphi}^2(t,\vec{x}) - 
\langle \widehat{\varphi}^2(t,\vec{x}) \rangle \Bigr)^2 \Biggr\rangle
= \int_0^{\infty} \!\!\! dZ \, \frac{(Z \!-\! \sigma^2)^2}{\sqrt{2 \pi
\sigma^2 Z}} \, \exp\Bigl[-\frac{Z}{2 \sigma^2}\Bigr] = 2 \sigma^4(t) \; .
\end{equation}
However, this only implies that that stochastic fluctuations about the 
mean are significant, even when considered as a fraction of the mean. 

What the expectation value $\langle \widehat{\varphi}^2(t,\vec{x}) \rangle
= \sigma^2(t)$ really tells us is:
\begin{itemize}
\item{That $\widehat{\varphi}^2(t,\vec{x})$ grows without bound; and}
\item{That this growth is proportional to the number of e-foldings, 
$\ln[a(t)]$.}
\end{itemize}
We can gain a quantitative assessment of the reliability of the
first conclusion by using (\ref{chisq}) to compute the probability that 
$\widehat{\varphi}^2(t,\vec{x})$ remains less that some constant
value $\Phi^2$,
\begin{eqnarray}
{\rm Prob}\Bigl( \widehat{\varphi}^2(t,\vec{x}) < \Phi^2\Bigr) & = &
\int_0^{\Phi^2} \!\!\! dZ \, \frac1{\sqrt{2 \pi \sigma^2(t) Z}} \,
\exp\Bigl[ -\frac{Z}{2 \sigma^2(t)} \Bigr] \; , \qquad \\
& = & \sqrt{\frac{ 2 \Phi^2}{\pi \sigma^2(t)}} \Biggl\{1 + 
O\Bigl( \frac{\Phi^2}{\sigma^2(t)} \Bigr) \Biggr\} \; .
\end{eqnarray}
Because $\sigma^2(t)$ grows like the number of e-foldings, we see that
the probability for $\widehat{\varphi}^2(t,\vec{x})$ to fall below any
fixed value $\Phi^2$ goes to zero at late times. Of course that 
vindicates the inference of growth without bound. The second inference 
can be tested by computing the probability for $\widehat{\varphi}^2(t,
\vec{x})$ to be above some time dependent value $\Phi^2(t)$ which grows 
faster than $\sigma^2(t)$,
\begin{eqnarray}
{\rm Prob}\Bigl( \widehat{\varphi}^2(t,\vec{x}) > \Phi^2(t)\Bigr) & = &
\int_{\Phi^2}^{\infty} \!\!\! dZ \, \frac1{\sqrt{2 \pi \sigma^2(t) Z}} \,
\exp\Bigl[ -\frac{Z}{2 \sigma^2(t)} \Bigr] \; , \qquad \\
& = & \sqrt{\frac{ 2 \sigma^2(t)}{\pi \Phi^2(t)}} \exp\Bigl[-
\frac{\Phi^2(t)}{2 \sigma^2(t)} \Bigr] \Biggl\{1 + 
O\Bigl( \frac{\sigma^2(t)}{\Phi^2(t)} \Bigr) \Biggr\} \; . \qquad
\end{eqnarray}
Under the assumption that $\sigma^2(t)/\Phi^2(t)$ goes to zero we see
that this probability also approaches zero. Hence the second inference
is equally valid, and we conclude that no serious error arises from
using expectation values to study $\widehat{\varphi}^2(t,\vec{x})$.

One might object that there is still a substantial disagreement between
quantum field theoretic expectation values and stochastic samples because
the former contain an ultraviolet divergent constant in addition to the
infrared logarithm \cite{old},
\begin{equation}
\Bigl\langle \Omega \Bigl\vert \varphi^2(t,\vec{x}) \Bigr\vert \Omega
\Bigr\rangle = \Bigl({\rm Divergent\ Constant}\Bigr) + \frac{H^2}{4 \pi^2}
\, \ln[a(t)] \; .
\end{equation}
The form of this divergence depends upon the regularization technique; with
dimensional regularization in $D$ spacetime dimensions one finds \cite{TWJMPW},
\begin{eqnarray}
\lefteqn{\Bigl\langle \Omega \Bigl\vert \varphi^2(t,\vec{x}) \Bigr\vert 
\Omega \Bigr\rangle = 
\frac{H^{D-2}}{ (4 \pi)^{\frac{D}2}} \frac{\Gamma(D \!-\!1)}{
\Gamma(\frac{D}2)} \Biggl\{ 2 \ln[a(t)]
-\psi\Bigl(1 \!-\! \frac{D}2\Bigr) } \nonumber \\
& & \hspace{4cm} + \psi\Bigl( \frac{D \!-\!1}2\Bigr) + \psi(D \!-\!1)
+ \psi(1) + O\Bigl(a^{-2}\Bigr) \Biggr\} . \qquad \label{fcoin}
\end{eqnarray}
However, many fully dimensionally regulated and renormalized computations 
have been done involving massless, minimally coupled scalars 
\cite{OW,PTW,PW,BOW,MW1,PTsW1} and gravitons \cite{TW1,MW2,KW} with various 
interactions. What always happens is that counterterms absorb the 
ultraviolet divergence and leave the infrared logarithm as the dominant
contribution to the final result. That is just what one gets from
$\langle \widehat{\varphi}^2(t,\vec{x}) \rangle$.

One might also object that the example of $\widehat{\varphi}^2(t,\vec{x})$ 
is contrived because it represents two fields at the same spacetime point.
However, it is exactly these terms which are most responsible for the growth 
of the vacuum energy in $\lambda \varphi^4$ theory \cite{OW} and for the 
photon developing a mass in scalar quantum electrodynamics \cite{four,PTW}. 
Note also that the use of expectation values gives precisely the correct 
results for the leading logarithm effects in each case \cite{OW,four,PTW},
\begin{eqnarray}
\Bigl\langle \Omega \Bigl\vert T_{\mu\nu}(t,\vec{x}) \Bigr\vert \Omega 
\Bigr\rangle & \longrightarrow & -\frac{\lambda}{4!} \Bigl\langle 
\widehat{\varphi}^4(t,\vec{x}) \Bigr\rangle g_{\mu\nu} = -\frac{\lambda}8
\Bigl[ \frac{H^2}{4 \pi^2} \, \ln[a(t)]\Bigr]^2 \, g_{\mu\nu} \; , \\
M^2_{\gamma} & \longrightarrow & +e^2 \Bigl\langle \widehat{\varphi}^2(t,
\vec{x}) \Bigr\rangle = \frac{e^2 H^2}{4 \pi^2} \, \ln[a(t)] \; .
\end{eqnarray}
This is not an accident, nor is the coincidence restricted to lowest order 
perturbative results such as those given above. For scalar potential models 
one can show that Starobinsky's formalism \cite{AAS} captures the leading 
infrared logarithms of quantum field theoretic expectation values to all 
orders \cite{TW2}. The best way of viewing the stochastic formalism is not 
as an alternative to expectation values but rather as marvelously simple 
way of deriving the most important contributions to them.

In some cases the stochastic formalism can do even more. Starobinsky and 
Yokoyama have shown how it can be used to sum the series of leading infrared 
logarithms to derive nonperturbative results for the late time limit 
\cite{SY}. For example, these results explicitly disprove the two simplest 
implementations of the common notion that cosmological evolution can be
viewed as a renormalization group flow \cite{RPW}. The Starobinsky-Yokoyama 
technique has recently been extended to Yukawa theory \cite{MW1} and to 
scalar quantum electrodynamics \cite{PTsW2}. It has not yet been extended 
to gravity but there are reasons for believing that some version of it can 
be \cite{MW3,TW3}.

\section{Fluctuations in Space}

We can study spatial variation by taking the difference of the fields
at points on the surface of simultaneity,
\begin{eqnarray}
\lefteqn{ \Delta \widehat{\varphi}(t,\Delta \vec{x}) \equiv 
\widehat{\varphi}(t,\vec{0}) - \widehat{\varphi}(t,\Delta \vec{x}) \; ,} \\
& & \hspace{-.5cm} = \! \sum_{\vec{n} \neq 0} \! \sqrt{\frac{2 H^5}{k^3}}
\, \theta\Bigl( H a(t) \!-\! k\Bigr) \sin\Bigl(\frac{\vec{k} \!\cdot\! 
\Delta \vec{x}}2 \Bigr) \Biggl[ \sin\Bigl( \frac{\vec{k} \!\cdot\! \Delta
\vec{x}}2 \Bigr) \, \alpha_{\vec{n}} + \cos\Bigl( \frac{\vec{k} \!\cdot\! 
\Delta \vec{x}}2 \Bigr) \, \beta_{\vec{n}} \Biggr] . \qquad
\end{eqnarray}
Just like $\widehat{\varphi}(t,\vec{x})$, this is a sum of independent
Gaussians, so $\Delta \widehat{\varphi}(t,\Delta \vec{x})$ is itself a
Gaussian. Because the mean of each constituent is zero, the mean of
$\Delta \widehat{\varphi}(t,\Delta \vec{x})$ also vanishes. Its variance
is the sum of the variance of each constituent,
\begin{eqnarray}
\lefteqn{\sigma^2(t,\Delta x) = \sum_{\vec{n} \neq 0} \frac{2 H^5}{k^3}
\, \theta\Bigl( H a(t) \!-\! k\Bigr) \sin^2\Bigl(\frac{\vec{k} \!\cdot\! 
\Delta \vec{x}}2 \Bigr) \; , } \\
& & \hspace{-.5cm} = 4\pi \int_{1/2\pi}^{a/2\pi} \!\!\!\!\!\!\!\! dn \, n^2 \, 
\frac{H^5}{(2 \pi H n)^3} \Bigl[1 - \frac{\sin(k \Delta x)}{k \Delta x}
\Bigr] \; , \\
& & \hspace{-.5cm} = \frac{H^2}{2 \pi^2} \int_{H \Delta x}^{a H \Delta x} 
\!\!\!\!\!\!\!\!  dz \, \Bigl[ \frac1{z} - \frac{\sin(z)}{z}\Bigr] \; , \\
& & \hspace{-.5cm} = \frac{H^2}{2 \pi^2} \Biggl\{ 
\frac{\sin[a(t) H \Delta x]}{a(t) H \Delta x} \!-\! 
\frac{\sin[H \Delta x]}{H \Delta x} \!+\! \ln[a(t)] - {\rm ci}[a(t) H \Delta x]
\!+\! {\rm ci}[H \Delta x] \Biggr\} . \qquad \label{var}
\end{eqnarray}
The symbol ${\rm ci}(x)$ stands for the cosine integral whose definition
and expansion for small $x$ are,
\begin{eqnarray}
{\rm ci}(x) & \equiv & -\int_x^{\infty} \!\!\! dt \, \frac{\cos(t)}{t} 
= \ln(x) + \gamma + \int_0^x \!\! dt \, \Bigl[ \frac{\cos(t) \!-\! 1}{t}
\Bigr] \; , \qquad \\
& = & \ln(x) + \gamma + \sum_{n=1}^{\infty} \frac{(-1)^2 x^{2n}}{2n \!
\cdot \! 2n!} \; . \label{small}
\end{eqnarray}
Note the appearance of Euler's constant, $\gamma \approx 0.577215665$.

Quantum gravitational perturbation theory breaks down when the expectation
value of $G \widehat{\varphi}^2(t,\vec{x})$ grows to be of order one 
\cite{TW3},
\begin{equation}
\Bigl\langle G \widehat{\varphi}^2(t,\vec{x}) \Bigr\rangle = 
\frac{G H^2}{4 \pi^2} \times \ln[a(t)] \sim 1 \quad \Longrightarrow \quad
\ln[a(t)] \sim \frac1{G H^2} \gg 1 \; .
\end{equation}
One might expect that this is also when back-reaction becomes significant.
At this point the fluctuation of each of the two fields in $\Delta 
\widehat{\varphi}(t,\Delta \vec{x})$ is enormous, as one can see from
expression (\ref{sig1}). Cosmologists frequently express the worry that,
if back-reaction of this sort ever becomes significant, it must induce
similarly large fluctuations in $\Delta \widehat{\varphi}(t,\Delta \vec{x})$.
That would lead to an unacceptable level of inhomogeneity in the 
post-inflationary universe.

We can test the cosmologists' fear by choosing $\Delta x$ so as to keep 
the physical distance a constant fraction $K$ of the Hubble length,
\begin{equation}
a(t) \Delta x = \frac{K}{H} \; .
\end{equation}
From expression (\ref{var}) and the asymptotic expansion (\ref{small}) we
see that the variance rapidly approaches a not especially large constant,
\begin{eqnarray}
\sigma^2(t,Ka^{-1}) & \!\!\!=\!\!\! & \frac{H^2}{2\pi^2} \Biggl\{ 
\frac{\sin(K)}{K} \!-\! \frac{\sin(K a^{-1})}{K a^{-1}} \!+\! \ln(a) \!-\! 
{\rm ci}(K) \!+\! {\rm ci}(K a^{-1}) \Biggr\} , \qquad \\
& \!\!\! = \!\!\! & \frac{H^2}{2 \pi^2} \Biggl\{ \frac{\sin(K)}{K} - 1 + 
\ln(K) + \gamma - {\rm ci}(K) + O(a^{-2}) \Biggr\} . \label{actual}
\end{eqnarray}
This should be compared with the variance of $\sigma^2(t) = H^2/4\pi^2 \times
\ln[a(t)]$ of each field in the difference. So it is false that spatial 
fluctuations between nearby points become enormous whenever back-reaction 
is significant.

Although our result (\ref{actual}) might seem surprising, there is a very 
simple and general reason for it based upon causality. The stochastic field 
acquires its fluctuations one instant at a time, as each new complement of 
modes experiences horizon crossing and contributes to the stochastic jitter. 
The result at any spacetime point $(t,\vec{x})$ depends upon what happened 
in the past light-cone of that point. After a long period of inflation, two 
fields a fixed distance apart very largely share the same past light-cone, 
so they experience almost the same fluctuations. This remains true even if 
the faction $K$ is much greater than one, because significant back-reaction
requires the staggering number of $1/GH^2 \sim 10^6$ e-foldings.

Let us also note that the level (\ref{actual}) of inhomogeneity we see 
is about right to explain the primordial power spectra \cite{TW3}. So far 
from the stochastic formalism invalidating theories of inflation which
are based on back-reaction, it provides an essential ingredient.

\section{Epilogue}

We have employed a very simple scalar model on de Sitter background
to examine the chief criticisms against using expectation values and 
in-out matrix elements to infer back-reaction:
\begin{enumerate}
\item{That stochastic fluctuations about the mean value change
the entire picture; and}
\item{That significant back-reaction necessarily implies an 
unacceptable level of inhomogeneity.}
\end{enumerate}
Neither criticism is supported by our study. In section 3 we found
that although there is significant stochastic fluctuation in the
quantity $\widehat{\varphi}^2(t,\vec{x})$, there is zero probability
either for this quantity to remain bounded, or for it to grow at a 
faster rate than that predicted by its mean value. In section 4 we 
showed that there is no growth in the variance of the difference between 
two stochastic fields held at a fixed physical distance from one another.

The cosmologists' objection that expectation values give a misleading
picture for some operators is certainly correct for 
$\widehat{\varphi}(t,\vec{x})$. On the other hand, expectation values
give a fair representation of $\widehat{\varphi}^2(t,\vec{x})$. So the 
picture that emerges is more complex than either a total refutation of 
expectation values or their complete vindication. The fair conclusion 
would seem to be that the sensitivity of a particular mechanism of 
back-reaction to stochastic fluctuations should always be checked, but 
there are no grounds for rejecting the mechanism prior to such a check.

We do not view the stochastic formalism of Starobinsky \cite{AAS} as an 
alternative to quantum field theory but rather as a wonderful tool
for isolating the most significant corrections \cite{TW2}. The really
intriguing thing is that the stochastic formalism can sometimes be
used to obtain nonperturbative results \cite{SY,MW1,PTsW2}. We believe
this can be done for quantum gravity \cite{MW3,TW3} and that it is 
worthwhile attempting to anticipate the result \cite{TW4}.

\centerline{\bf Acknowledgements}
We are grateful to S. Habib and A. Linde for stimulating discussions 
at the inception of this project. This work was partially supported 
by the European Union grant FP-7-REGPOT-2008-1-CreteHEPCosmo-228644,
by National Science Foundation Grants PHY-0653085 and PHY-0855021, 
and by the Institute for Fundamental Theory at the University of Florida.

\end{document}